\documentclass{PoS}

\usepackage{graphicx} 

\title{Observation of Radiative $B^0 \rightarrow \phi K^0 \gamma$ Decays and Measurements of Their Time-Dependent $CP$ Violation}

\ShortTitle{Observation of $B^0 \rightarrow \phi K^0 \gamma$ and Measurements of $CP$ Violation}

\author{\speaker{Himansu Sahoo}\thanks{on behalf of the Belle Collaboration.}\\
        Department of Physics and Astronomy,\\
        University of Hawaii,\\ 
        Honolulu, HI 96822, USA\\
        E-mail: \email{himansu@phys.hawaii.edu}}

\newcommand{\btopkg}{B \rightarrow \phi K \gamma}
\newcommand{\btopkog}{B^0 \rightarrow \phi K^0 \gamma}
\newcommand{\btopksg}{B^0 \rightarrow \phi K_S^0 \gamma}
\newcommand{\btopkpg}{B^+ \rightarrow \phi K^+ \gamma}

\abstract{
We report the first observation of the radiative decay $\btopkog$ 
using a data sample 
of $772 \times 10^6$ $B\overline{B}$ pairs collected at 
the $\Upsilon(4S)$ resonance 
with the Belle detector at the KEKB asymmetric-energy $e^+e^-$ collider. 
We observe a signal of 
$35\pm8$ 
events with a significance of 
$5.4$ standard deviations
including systematic uncertainties. The measured branching fraction is 
${\cal B}(\btopkog) = (2.66\pm 0.60 \pm 0.32) \times 10^{-6}$, where
the uncertainties are statistical and systematic, respectively.
We also report the first measurement of time-dependent 
$CP$ violation parameters:
${\mathcal S}_{\phi K_S^0 \gamma} = +0.74^{+0.72}_{-1.05} (\rm{stat})^{+0.10}_{-0.24} (\rm{syst})$ and
${\mathcal A}_{\phi K_S^0 \gamma} = +0.35 \pm 0.58 (\rm{stat})^{+0.23}_{-0.10} (\rm{syst})$.
We also precisely measure
${\mathcal B}(\btopkpg) = (2.34\pm 0.29 \pm 0.23) \times 10^{-6}$.
The observed $M_{\phi K}$ mass spectrum differs
significantly from that expected in a three-body phase-space decay.
These results are preliminary.
}

\FullConference{35th International Conference of High Energy Physics - ICHEP2010,\\
		July 22-28, 2010\\
		Paris France}

\begin{document}


Rare radiative $B$ meson decays play an important role in the 
search for physics 
beyond the standard model (SM).
These decays are forbidden at tree level in the SM, but allowed through
electroweak loop processes. The loop can be mediated by non-SM particles,
and therefore is sensitive to new physics (NP).
Exclusive $b \rightarrow s \gamma$ decays have 
been extensively measured,
but their total sum so far accounts only for $44\%$ of the inclusive rate.
Therefore, further measurements of branching fractions for exclusive 
$\btopkg$ modes will improve our understanding of 
the $b \rightarrow s \gamma$ process. 
The emitted photons are predominantly left-handed (right-handed) 
in $b \rightarrow s \gamma$ 
($\overline{b} \rightarrow \overline{s} \gamma$) decays. 
This suppresses the 
$CP$ asymmetry in the SM by the quark mass ratio ($2m_s/m_b$).
The expected mixing-induced $CP$ asymmetry parameter (${\cal S}$)
is ${\cal O}(3\%)$
and the direct $CP$ asymmetry parameter 
(${\cal A}$) is $\sim 0.6\%$~\cite{ags}.
However, in several extensions of the SM, both the photon helicities might
contribute to the decay.
Therefore, any significantly larger $CP$ asymmetry 
would be clear evidence for NP.

In this presentation, we report the first observation of 
neutral mode $\btopkog$~\cite{conj}, 
the first measurements of time-dependent $CP$ violation 
in this mode, as well as improved measurements 
in $\btopkpg$ 
using a data sample of $772 \times 10^6$ $B\overline{B}$ pairs
collected at the $\Upsilon(4S)$ 
resonance with the Belle detector~\cite{belledetector}
at the KEKB asymmetric-energy $e^+e^-$ (3.5 on 8.0 GeV) 
collider~\cite{kekb}.
This data sample is
nearly eight times larger than the sample used in 
our previous measurement~\cite{alex_prl}.
%
In the decay chain 
$\Upsilon(4S) \rightarrow B^0 \overline{B}{}^0 \rightarrow f_{\rm rec}f_{\rm tag}$, 
where one of the $B$ mesons decays at 
time $t_{\rm rec}$ to the signal mode $f_{\rm rec}$ and the other decays
at time $t_{\rm tag}$ to a final state $f_{\rm tag}$ that distinguishes
between $B^0$ and $\overline{B}{}^0$, the decay rate has a time 
dependence given by
\begin{eqnarray}
\label{eq_decay}
{\mathcal P}(\Delta{t})= \frac{ e^{-|\Delta{t}|/{\tau_{B^0}}} }{4\tau_{B^0}}
\biggl\{1 + q \,
 \Bigl[ {\mathcal S} \sin(\Delta m_d \Delta{t})
  +  {\mathcal A} \cos(\Delta m_d \Delta{t})
\Bigr] \biggr\}.
\end{eqnarray}
Here $\tau_{B^0}$ is the neutral $B$ lifetime,
$\Delta m_d$ is the mass difference between the two 
neutral $B$ mass eigenstates, 
$\Delta t = t_{\rm rec} - t_{\rm tag}$, and the $b$-flavor charge 
$q$ equals $+1$ $(-1)$ when the tagging $B$ meson is a 
$B^0$ ($\overline{B}{}^0$). 


Signal candidates are reconstructed in the 
$\btopkpg$ and $\btopksg$ modes, with 
$\phi \rightarrow K^+K^-$ and $K_S^0 \rightarrow \pi^+\pi^-$.
The invariant mass of the kaon pairs from $\phi$ is required to be within
$\left|M_{K^+K^-}-m_{\phi}\right| < 0.01$ GeV/$c^2$, 
where $m_{\phi}$ denotes the $\phi$ meson world-average mass~\cite{pdg}.
The $K_S^0$ selection criteria are the same as those 
described in the Ref.~\cite{belle_b2s}; the invariant mass of the pion
pairs is required to satisfy
$M_{\pi^+\pi^-} \in (0.482, 0.514)$ GeV/$c^2$.
The high energy prompt photons are selected from isolated clusters 
within the barrel region of the calorimeter
and center-of-mass system (cms) energy, 
$E_{\gamma}^{\rm cms} \in (1.4,3.4)$ GeV.
$B$ candidates are identified using 
two kinematic variables: the energy difference 
$\Delta E \equiv E_B^{\rm cms} - E_{\rm beam}^{\rm cms}$ and the
beam-energy-constrained mass 
$M_{\rm bc} \equiv \sqrt{(E_{\rm beam}^{\rm cms})^2 - (p_B^{\rm cms})^2}$,
where $E_{\rm beam}^{\rm cms}$ is the beam energy in the cms, and 
$E_B^{\rm cms}$ and $p_B^{\rm cms}$ are the cms energy and momentum, 
respectively, of the reconstructed $B$ candidate.
In the $M_{\rm bc}$ calculation, the photon momentum is replaced by
$(E_{\rm beam}^{\rm cms} - E_{\phi K}^{\rm cms})$ 
to improve its resolution.
The candidates
that satisfy the requirements
$M_{\rm bc} > 5.2 \;{\rm GeV/}c^2$ and 
$\left|\Delta E\right| < 0.3 \;\rm{GeV}$ 
are selected for further analysis.

\begin{figure}[htbp]
\begin{center}
\includegraphics[width=0.24\columnwidth]{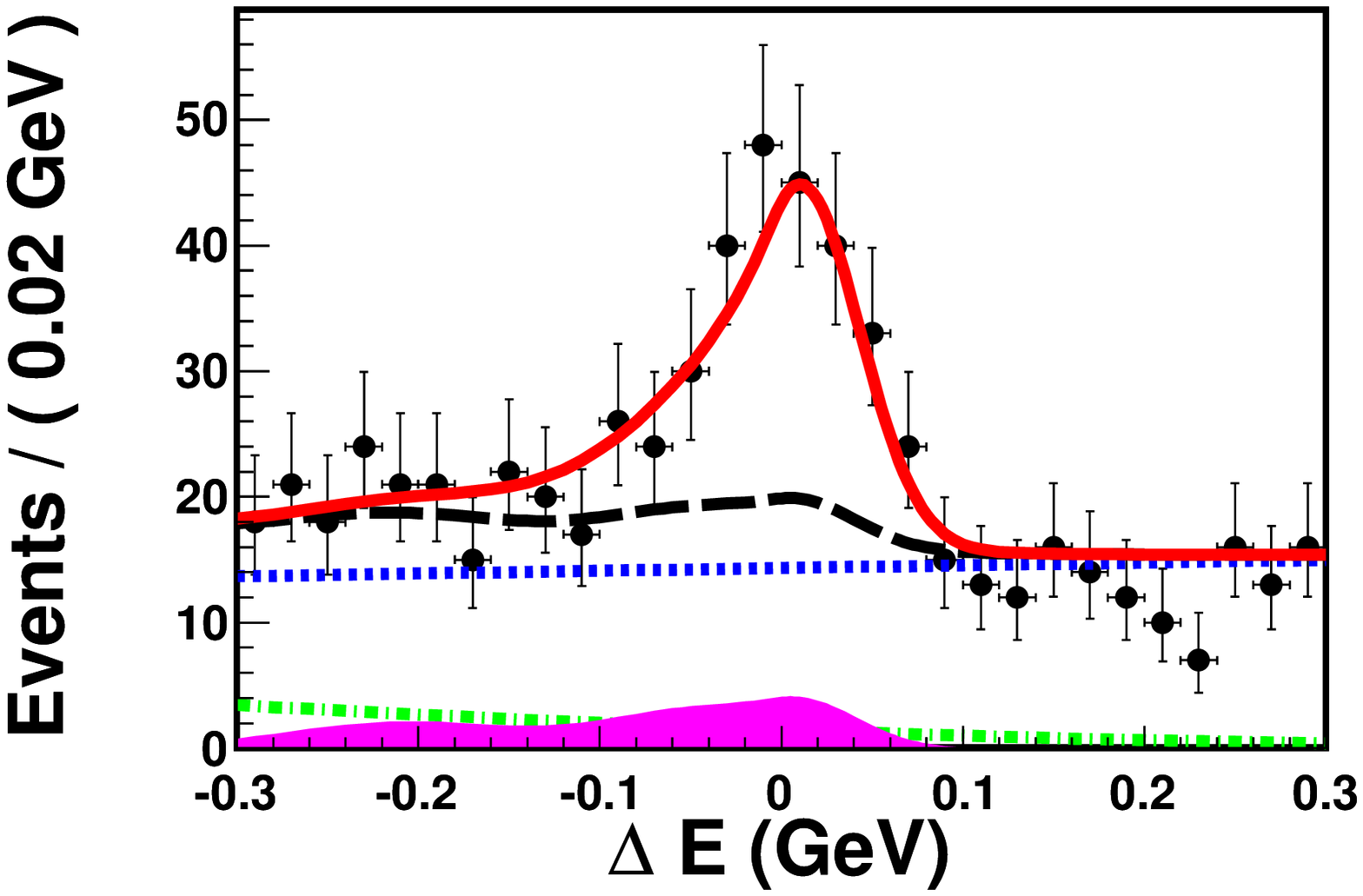}
\includegraphics[width=0.24\columnwidth]{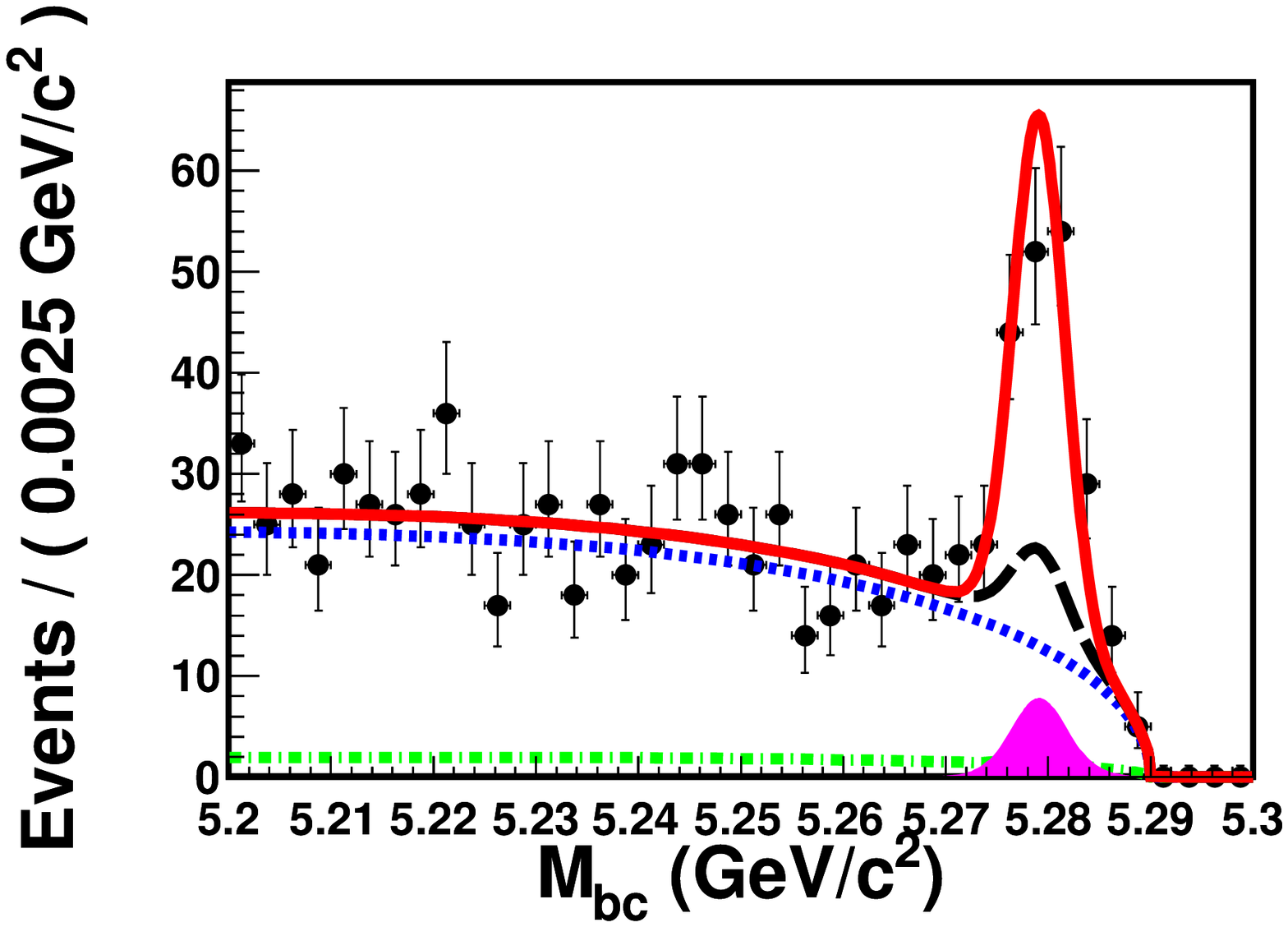}
\includegraphics[width=0.24\columnwidth]{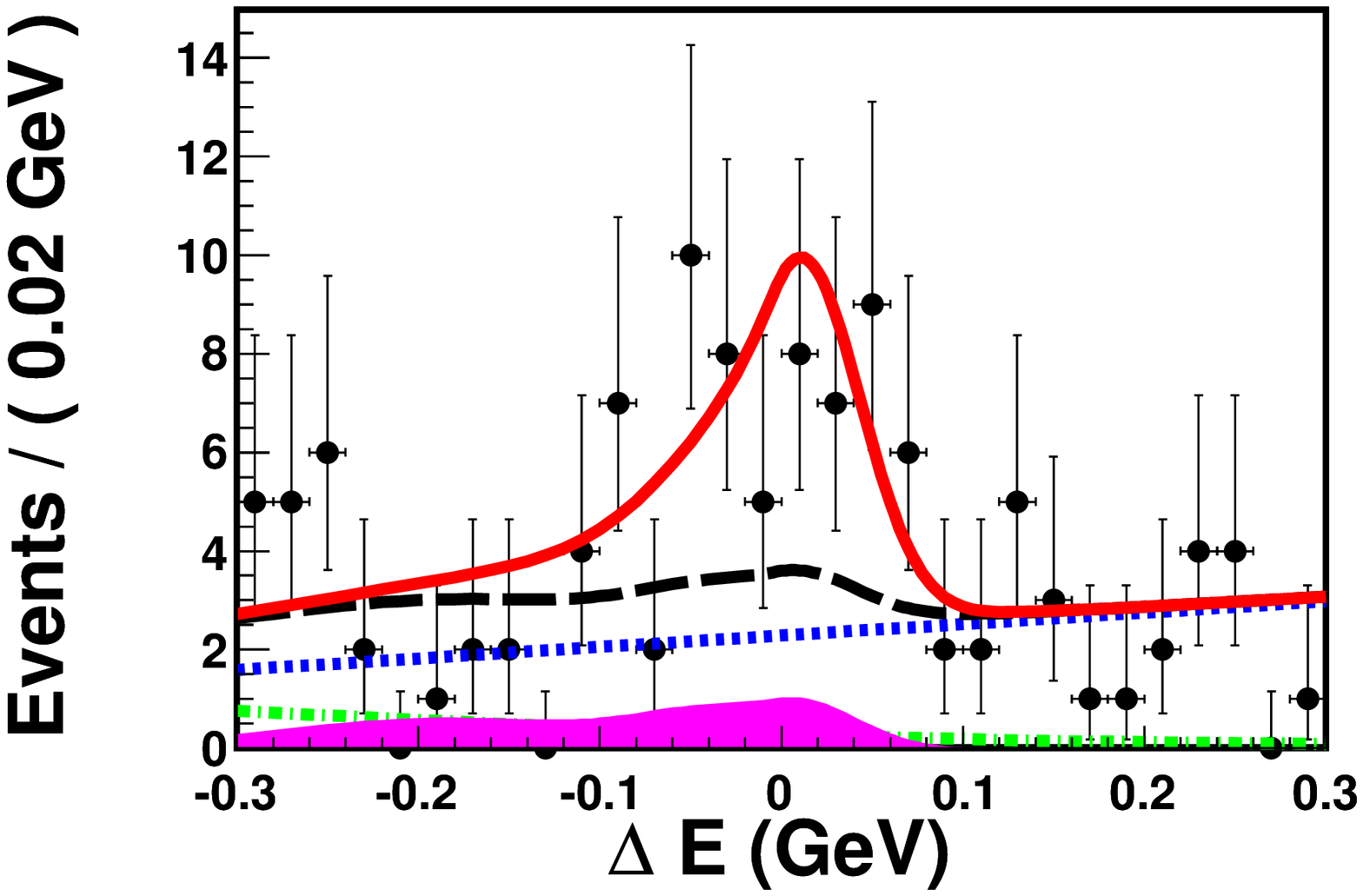}
\includegraphics[width=0.24\columnwidth]{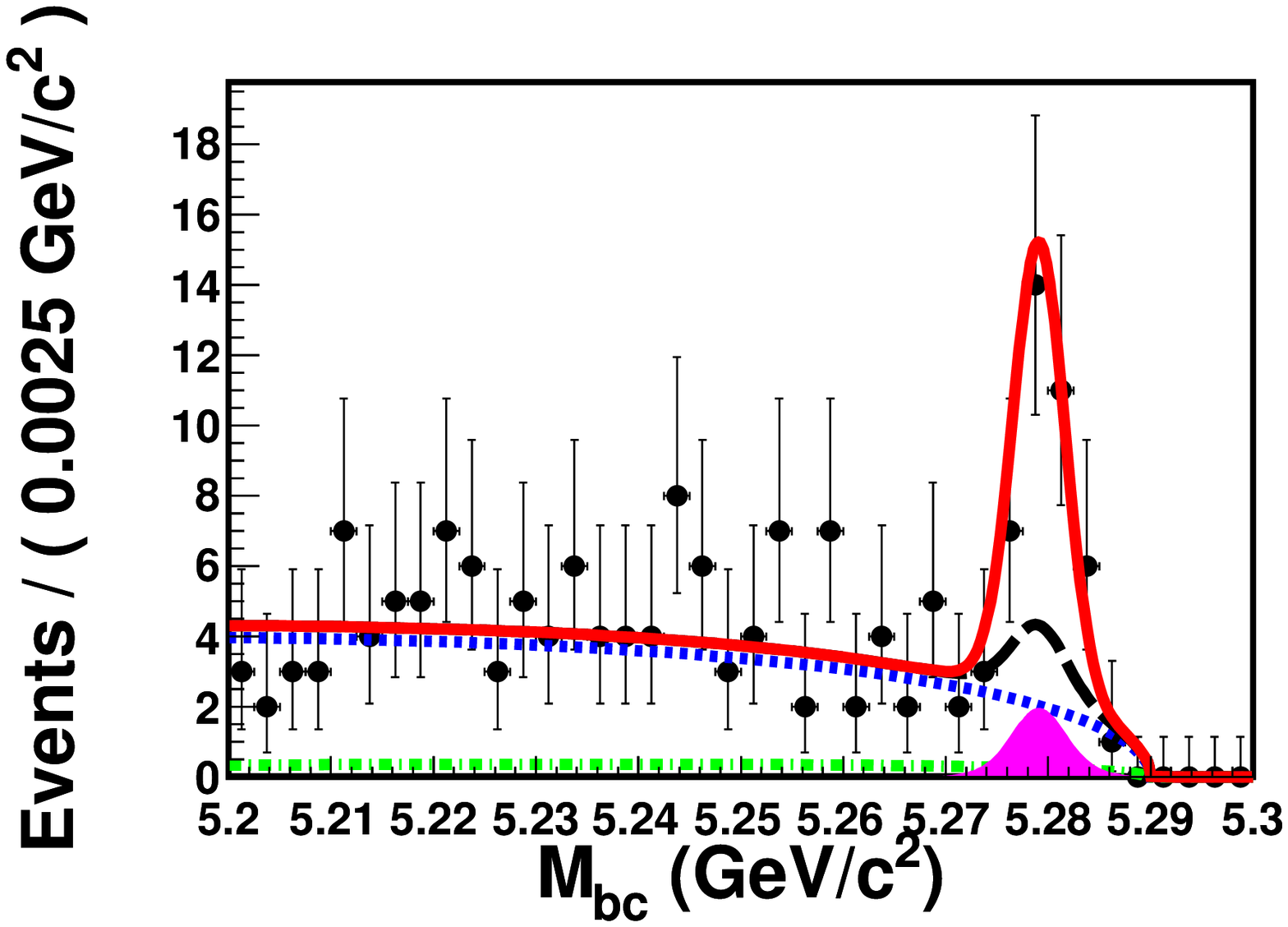}
 \put(-393,47){\bf (a)}
 \put(-280,47){\bf (b)}
 \put(-185,47){ \bf (c)}
 \put(-70,47){\bf (d)}
\end{center}
\vspace{-0.8cm}
\caption{The $\Delta E$ and $M_{\rm bc}$ projections 
for $\btopkpg$ ((a) and (b)) and $\btopksg$ ((c) and (d)). 
The points with error bars represent the data. The curves show the total fit function (solid red), total background function (long-dashed black), continuum component (dotted blue), the $b\rightarrow c$ component (dashed-dotted green) and the non-resonant component as well as other charmless backgrounds (filled magenta histogram).
}
\label{fig:dembc}
\end{figure}

The dominant background is from the continuum process, 
which is suppressed by a requirement on likelihood ratio ($\mathcal{R}$) 
from event shape variables and the $B$ flight direction.
We require $\mathcal{R}>0.65$, which removes $91\%$ 
of the continuum while retaining $76\%$ of the signal.
In the $\btopksg$ mode,
some backgrounds from $b \rightarrow c$ decays, such as
$D^0\pi^0$, $D^0\eta$ and $D^-\rho^+$
peak in the $M_{\rm bc}$ distribution. 
We remove the dominant peaking backgrounds by vetoing
$\phi K_S^0$ combinations consistent with the 
nominal $D$ mass~\cite{pdg}.
Some of the charmless backgrounds, where the $B$ meson decays to
$\phi K^{*}(892)$, $\phi K \pi^0$ and $\phi K \eta$ 
also peak in $M_{\rm bc}$, but are shifted towards lower $\Delta E$. 
Another significant background is
non-resonant $B \rightarrow K^+ K^- K \gamma$, which peaks in the
$\Delta E$-$M_{\rm bc}$ signal region; it is estimated using
the $\phi$ mass sideband
$M_{K^+K^-} \in (1.05,1.3)$ GeV/$c^2$, in data.

The signal yield is obtained from an extended
unbinned maximum-likelihood (UML) fit to the two-dimensional 
$\Delta E$-$M_{\rm bc}$ distribution as shown in Fig.~\ref{fig:dembc}.
The probability density functions (PDFs) for different components 
and fitting procedure are described 
in detail elsewhere~\cite{sahoo}.
The fit yields a signal of 
$136\pm17$ $\btopkpg$ and $35\pm8$ $\btopksg$ candidates.
The signal in the charged mode has a significance of
$9.6\,\sigma$, whereas that for the neutral mode is $5.4\,\sigma$,
including systematic uncertainties.
%
The observed $\phi K$ mass spectrum differs 
significantly from that expected in a 
three-body phase-space decay. 
Nearly $72\%$ of the signal events are concentrated
in the low-mass region 
($M_{\phi K} \in (1.5,2.0)$ GeV/$c^2$). 
The Monte Carlo (MC) efficiency after reweighting according to 
this $M_{\phi K}$ dependence is
$(15.3\pm0.1)\%$ for the charged and 
$(10.0\pm0.1)\%$ for the neutral mode.
We measure the 
branching fractions as 
${\mathcal B}(\btopkpg) = (2.34\pm 0.29 \pm 0.23) \times 10^{-6}$ and
${\mathcal B}(\btopkog) = (2.66\pm 0.60 \pm 0.32) \times 10^{-6}$,
where the uncertainties are statistical and systematic, respectively.



\begin{figure}[htbp]
\vspace{-0.2cm}
\begin{center}
\resizebox{0.4\columnwidth}{!}{\includegraphics{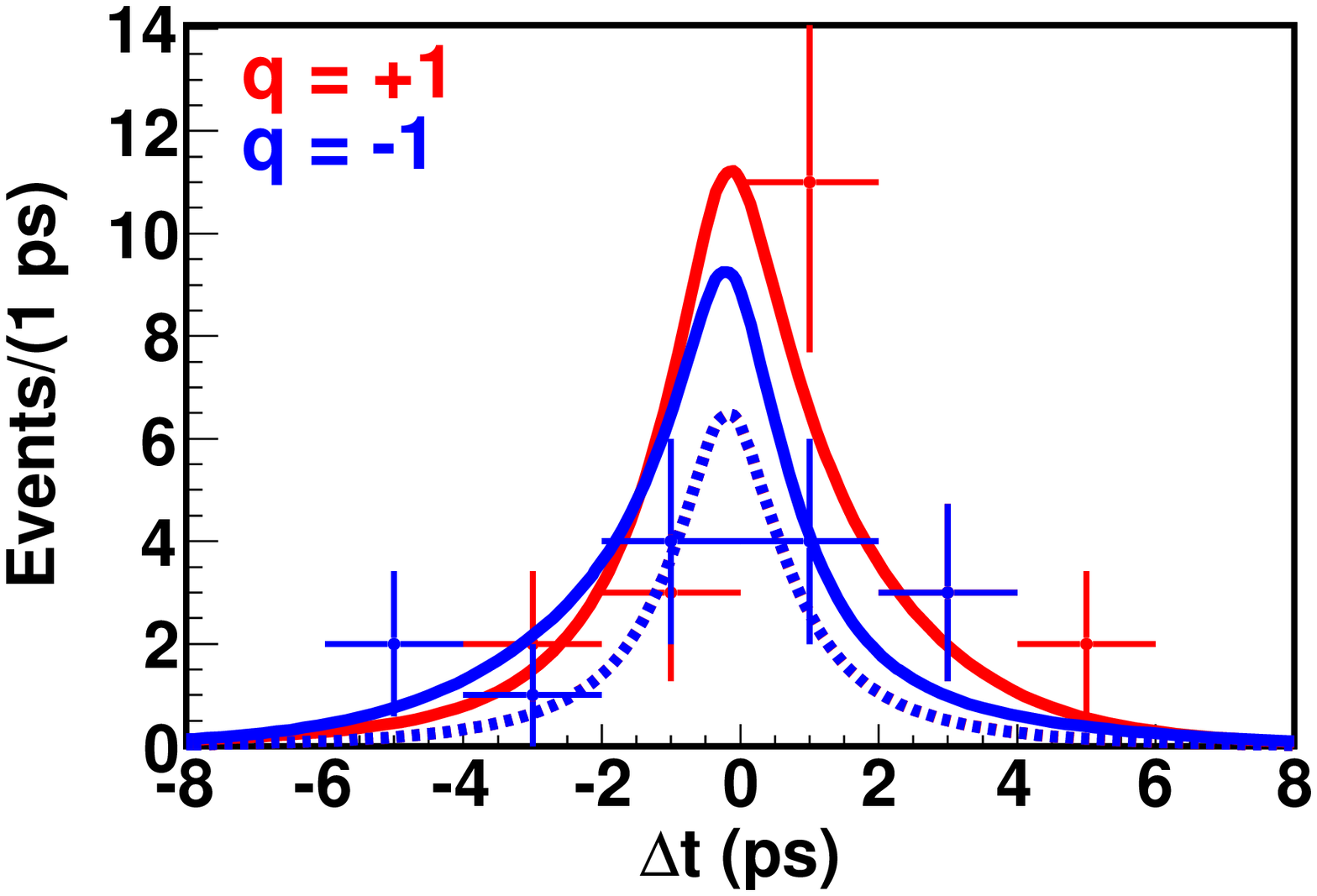}}
\resizebox{0.4\columnwidth}{!}{\includegraphics{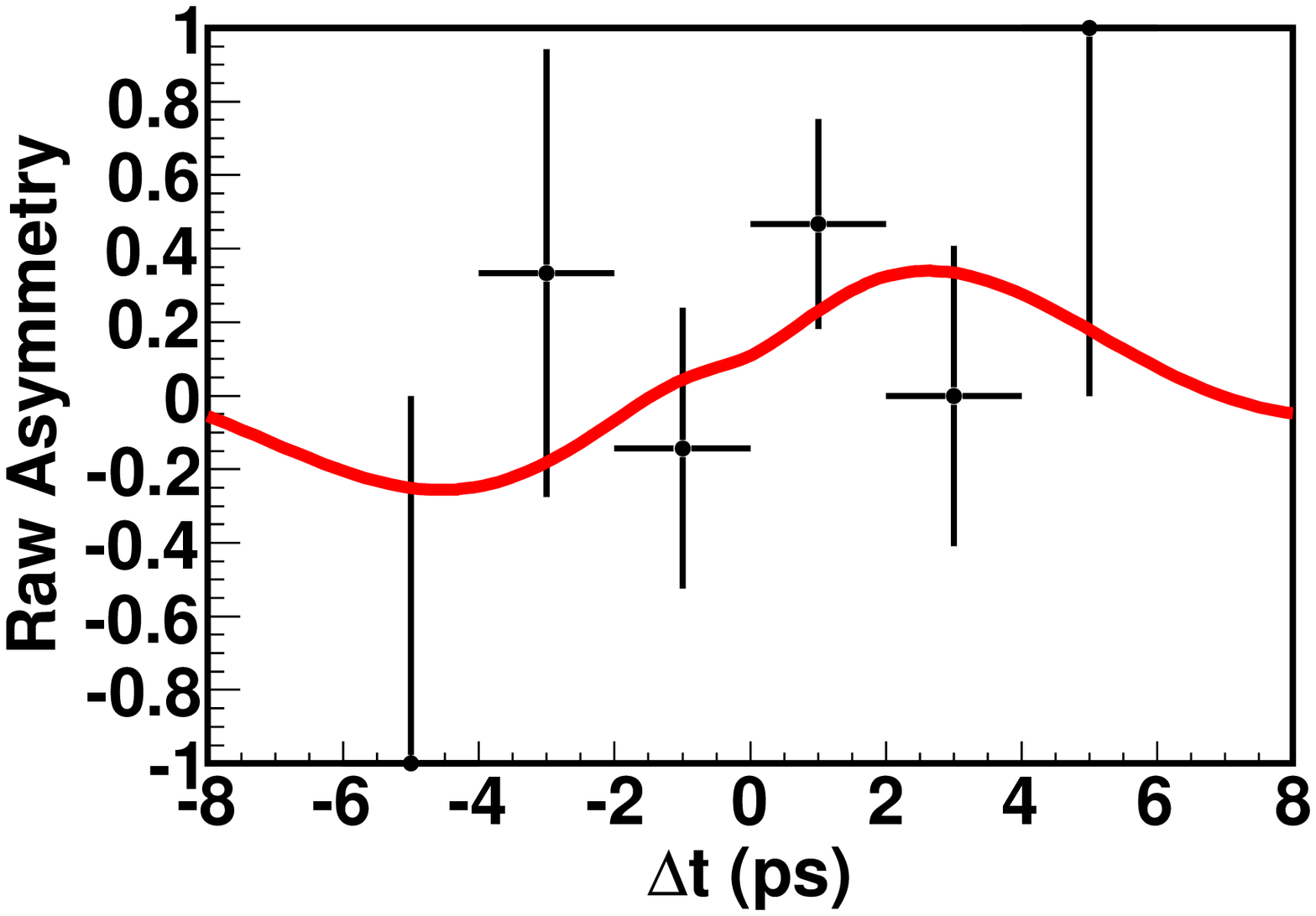}}
\end{center}
\vspace{-0.8cm}
\caption{$\Delta t$ distributions for $q$ = $+1$ and $q$ = $-1$ (left)
and the raw asymmetry (right) for well-tagged events. 
The dashed curves in the $\Delta t$ plot are the sum of 
backgrounds while the solid curves are the sum of signal and
backgrounds. The solid curve in asymmetry plot shows the result
of the UML fit.}
\label{fig:raw_ass}
\end{figure}

For the $CP$ asymmetry fit, we select events 
that satisfy
$M_{\rm bc} \in (5.27,5.29)$ GeV/$c^2$ and 
$\Delta E \in (-0.2,0.1)$ GeV.
The vertex position for the $f_{\rm rec}$ decay 
is reconstructed using the two kaon tracks from the $\phi$ meson
and that of $f_{\rm tag}$ decay is obtained from well-reconstructed tracks
that are not assigned to $f_{\rm rec}$~\cite{tag}. 
We then use a flavor tagging algorithm~\cite{tajima} 
to obtain $q$ and the
tagging quality factor $r \in (0,1)$.
The typical vertex reconstruction efficiency ($z$ resolution) is
$96\%$ ($115\,\mu$m) for $f_{\rm rec}$ and 
$94\%$ ($104\,\mu$m) for $f_{\rm tag}$.
After all selection criteria are applied, 
we obtain $75$ ($436$) events for the $CP$ fit 
with a purity of $45\%$ ($35\%$) in the neutral (charged) mode.
The signal PDF
is given by a modified form of Eq.~(\ref{eq_decay}) 
after fixing $\tau_{B^0}$ and $\Delta m_d$ to their 
world-average values~\cite{pdg}
and incorporating the effect of incorrect flavor assignment.
Since the non-resonant component has the same NP as the signal 
$B \to \phi K \gamma$, we treat this as signal.
The only free parameters in the $CP$ fit 
are ${\mathcal S}$ and ${\mathcal A}$, which are determined to be
${\mathcal S}_{\phi K_S^0\gamma} = +0.74^{+0.72}_{-1.05} (\rm stat)^{+0.10}_{-0.24} (\rm syst)$ and
${\mathcal A}_{\phi K_S^0\gamma} = +0.35 \pm 0.58 (\rm stat)^{+0.23}_{-0.10} (\rm syst)$.
We define the raw asymmetry in each $\Delta t$ bin by 
$(N_{+}-N_{-})/(N_{+}+N_{-})$, where $N_{+}$ $(N_{-})$
is the number of observed candidates with $q=+1$ $(-1)$.
Figure~\ref{fig:raw_ass} shows the $\Delta t$ distributions 
and raw asymmetry for events with good tagging quality ($r > 0.5$, $48\%$ of total).
We find that the error on ${\cal S}$ in the MINUIT minimization 
is much smaller than the expectation from MC 
simulation and has a probability of only $0.6\%$~\cite{minosvalue}.
This is due to low statistics and the presence of a single
special event. 
Instead, we use the $\pm 68\%$ confidence intervals 
in the residual distributions of ${\cal S}$ and ${\cal A}$, 
determined from the toy MC
as the uncertainties for the final result.

In summary, we report the first observation of radiative 
$\btopkog$ decays using a data sample of 
$772 \times 10^6$ $B\overline{B}$ pairs. The observed signal yield is
$35\pm8$ with a significance of $5.4\,\sigma$.
We also report the first measurements (preliminary) of time-dependent 
$CP$ violation parameters in this mode: 
${\mathcal S} = 0.74^{+0.72+0.10}_{-1.05-0.24}$ and
${\mathcal A} = 0.35\pm 0.58^{+0.23}_{-0.10}$.
We precisely measure the branching fraction
and charge asymmetry
(${\mathcal A}_{CP} = -0.03\pm 0.11\pm 0.08$) for the charged mode.
The signal events are mostly concentrated at low $\phi K$ mass,
which is similar to a two-body radiative decay.
With the present statistics, 
these measurements are consistent with the 
SM predictions and
there is no indication of NP from right-handed 
currents in radiative $B$ decays.
Much more luminosity is necessary for a 
precise test of the SM.
%

The author wish to thank the KEKB group for excellent operation of the
accelerator, the KEK cryogenics group for efficient solenoid
operations, and the KEK computer group and
the NII for valuable computing and SINET3 network support.  


\begin{thebibliography}{99}
\bibitem{conj}
Throughout this paper, the inclusion of the charge-conjugate 
decay mode is implied unless otherwise stated.
\bibitem{minosvalue}
The MINOS errors are $^{+0.32}_{-0.45}$ for ${\cal S}$
and $\pm 0.45$ for ${\cal A}$. The toy MC distribution has a width of
$^{+0.72}_{-1.05}$ for ${\cal S}$ and $\pm 0.58$ for ${\cal A}$.
\end{thebibliography}
\end{document}